\documentclass{EPS}

\usepackage{lineno}
\usepackage{graphicx}
\usepackage{color}
\usepackage{footmisc}

\title{Statistical properties of superflares on solar-type stars based on 1-min cadence data}
\author{
Hiroyuki Maehara, Okayama Astrophysical Observatory, National Astronomical Observatory of Japan,
3037-5 Honjo, Kamogata, Asakuchi,
Okayama 719-0232, Japan,
h.maehara@oao.nao.ac.jp\\
Takuya Shibayama, Nagoya University, shibayama@stelab.nagoya-u.ac.jp\\
Yuta Notsu, Kyoto University, ynotsu@kwasan.kyoto-u.ac.jp\\
Shota Notsu, Kyoto University, snotsu@kusastro.kyoto-u.ac.jp\\
Satoshi Honda, University of Hyogo, honda@nhao.jp\\
Daisaku Nogami, Kyoto University, nogami@kusastro.kyoto-u.ac.jp\\
Kazunari Shibata, Kyoto University, shibata@kwasan.kyoto-u.ac.jp
}

\abstract{
We searched for superflares on solar-type stars using Kepler data with
1 min sampling in order to detect superflares with short duration.
We found 187 superflares on 23 solar-type stars whose bolometric energy
ranges from the order of $10^{32}$ erg to $10^{36}$ erg.
Some superflares show multiple peaks with the peak 
separation of the order of $100$-$1000$ seconds which is comparable
to the periods of quasi-periodic pulsations in solar and stellar flares.
Using these new data combined with the results from the data with 30 min
sampling, we found the occurrence frequency ($dN/dE$) of superflares
as a function of flare energy ($E$) shows the power-law distribution
($dN/dE \propto E^{-\alpha}$) with $\alpha \sim -1.5$ for $10^{33}<E<10^{36}$
erg which is consistent with the previous results.
The average occurrence rate of superflares with the energy of $10^{33}$ erg 
which is equivalent to X100 solar flares is about once in 500-600 years.
The upper limit of energy released by superflares is 
basically comparable to a fraction of the
magnetic energy stored near starspots which is estimated 
from the photometry.
We also found that the duration of superflares ($\tau$) increases with the flare energy ($E$) 
as $\tau \propto E^{0.39\pm 0.03}$.
This can be explained if we assume the time-scale
of flares is determined by the Alfv$\acute{\rm e}$n time.}


\begin{document}

\maketitle

\section{Introduction}
 Solar flares are eruptive events in the solar atmosphere caused by
the magnetic reconnection \citep[e.g.][]{Shibata2011}.
The occurrence frequency of solar-flares  decreases as the flare energy increases.
The frequency-energy distribution of solar flares 
can be fitted by a simple power-law function with an index of $-1.5$ - $-1.9$
in the flare energy range between $10^{24}$ erg and $10^{32}$ erg
\citep[e.g.][]{Aschwanden2000, Shimizu1995, Crosby1993}.
The total bolometric energy released by the largest solar flares
is estimated to be the order of  $10^{32}$ erg \citep{Emslie2012} and the occurrence frequency of such flare is about once in 10 years.
On the other hand, 
much larger flares called "superflares" have been observed on
a wide variety of stars including solar-type stars \citep{Landini1986,Schaefer1989, Schaefer2000}.
Recently, many superflares on solar-type stars (G-type main sequence stars) 
have been discovered by the Kepler space telescope \citep{Maehara2012, Shibayama2013, Wu2015}.
The bolometric energy released by the superflare ranges from $10^{33}$ to 
$10^{36}$ erg which is $10$-$10^4$ times larger than that of the largest solar flares ($10^{32}$ erg).
Although most of G-type main sequence stars exhibiting superflares are rapidly rotating stars, 
the temperature and rotation period of some superflare stars are close to those of the Sun. 
For example, \cite{Nogami2014} reported that the rotational velocities of two superflare stars, 
KIC 9766237 and KIC 9944137, are comparable to that of our Sun and no hint of
binarity was detected. The average strength of the magnetic field on the surface of these stars,
by using the absorption line of Ca II at $\lambda$ 8542 \AA,  are estimated to be 1-20 G which
is comparable to or one order of magnitude higher than that of the Sun ($\sim 2$ G). 
On the other hand, \citep{Wichmann2014} reported that some superflare stars with
photometric period of 10-20 days have large rotational velocities of $v\sin i \sim$ 10-20 km/s.
This suggests that these stars are either sub-giant stars or binary systems, not main sequence
stars like our Sun.

Solar flares are sometimes accompanied by coronal mass ejections (CMEs).
Earth-directed CMEs can cause geomagnetic storms which can produce 
geomagnetically induced currents (GICs) in long electrical conductors 
(e.g. power lines) and can damage electrical equipment.
For example, on March 13, 1989 about 6 million people in Quebec, 
Canada suffered an electrical power blackout caused by a huge geomagnetic storm \citep{Allen1989}. 
Moreover, in 1859, one of the largest solar flares, called the Carrington flare
\citep{Carrington1859}, caused the largest geomagnetic
storm in the past 200 years \citep{Tsurutani2003} which led to a failure of telegraph systems in Europe and North America \citep{Loomis1861}.
Even the solar flares which release up to $10^{32}$ erg of energy can damage
not only our power grids, but also communication and navigation systems.
If superflares were to occur on the Sun, they would cause much more serious damage to our civilization.
Therefore it is important to study the properties of superflares
on solar-type stars.

Previous studies of superflares on the basis of the Kepler data mainly focus on 
superflares with extremely large energy ($> 10^{34}$ erg) and 
long duration ($> 1$ hour) and could not discuss the statistical properties
of superflares with energy between $10^{33}$ erg - $10^{34}$ erg due to
low time-resolution of the data.
In this paper, we report the results from the data with high time-resolution ($\sim 1$ min) and discuss the statistical properties (e.g. occurrence frequency, flare duration) of superflares on solar-type stars.

\section{Methods and Data}
We searched for flares (sudden brightenings of the star)
from the short-cadence data ($\sim 1$ min interval) observed with the 
Kepler space telescope between 2009 April (quarter 0: Q0) and 2013 May (Q17) \citep{Koch2010, Gilliland2010}.
We selected solar-type (G-type main sequence) stars from the data set
by using the surface temperature of the star ($T_{\rm eff}$) and the surface
gravity ($\log g$) 
taken from \cite{Huber2014} instead of those from \cite{Brown2011} (initial Kepler Input Catalog).
In previous papers \citep{Maehara2012, Shibayama2013}, 
we used stellar parameters taken from \cite{Brown2011}
and the selection criteria of $5100 \rm{K} < T_{\rm eff} < 6000$ K and $\log g >4.0 $.
However the temperatures in \cite{Brown2011} are
systematically lower by $\sim 200$ K than those in \cite{Huber2014}.
Since the occurrence frequency of superflares depends on the surface temperature \citep[e.g.][]{Candelaresi2014},
this systematic difference in surface temperature of stars would affect the resultant
flare occurrence frequency.
Therefore, in order to reduce the difference in the occurrence frequency of superflares
caused by the systematic difference in temperature, 
we used the selection criteria of $5300 \rm{K} < T_{\rm eff} < 6300 \rm{K}$
and $4.0 < \log g < 4.8$.
The total number of solar-type stars observed with short-cadence mode is
1547.

We used the Presearch Data Conditioned (PDC) light curve \citep{Stumpe2012, Smith2012}
for the detection of flares.
The typical length 
of each continuous observation in short-cadence mode is about 30 days.
We first calculated the distribution of the following parameter ($\Delta F^2$) for 
all consecutive data points from two pairs of the stellar brightness ($F(t)$) at 
the time of $t_{i-n-1}$, $t_{i-n}$, $t_{i}$, and $t_{i+1}$ ($i$ and $n$ are integer
numbers and $t_{i}$ means time of the $i$-th data point):
\begin{linenomath}
\begin{equation}
\Delta F^2(t_{i, n}) = s (F_{i}-F_{i-n-1})(F_{i+1}-F_{i-n}),\\
\end{equation}
\end{linenomath}
where
\begin{linenomath}
\begin{equation}
s=\left \{
\begin{array}{ll}
1  & (F_{i}-F_{i-n-1}>0 \ {\rm and}\  F_{i+1}-F_{i-n}>0) \\
-1  & (F_{i}-F_{i-n-1}<0 \ {\rm or}\  F_{i+1}-F_{i-n}<0)
\end{array}
\right.
\end{equation}
\end{linenomath}
In the case of typical 30-day short-cadence data blocks, $i$ ranges from 1 to $\sim 44000$.
If the flare peak is around at $t_{i}$ and the stellar brightness at 
$t_{i-n}$ is close to the quiescent level, 
the value of $\Delta F^2(t_{i})$ becomes much larger than 
the mode of the distribution.
Therefore $t_{i}-t_{i-n}$ should be comparable to or longer than 
the rise time of superflares.
According to \cite{Shibayama2013}, the rise time of superflare on
KIC 11610797 is about 2 min.
In the case of the X17  \footnotemark  solar flare \citep{Kopp2005}, the rise time is about
4 min.
Thus we used the separation between two data point $n=3$ ($t_{i}-t_{i-n} \sim 3$ min)
and $n=10$ ($\sim 10$ min)
for the analysis. In the case of larger $n$ value, we can detect flares having longer rise-time.
The threshold of the
flare was determined to be three times the value at the top 1\% of
the distribution.

\footnotetext{Solar flares are classified as A, B, C, M, or X, according to the peak X-ray (1-8 $\AA$) flux
 as measured by the GOES satellite. Peak fluxes of A, B, C, M and X class solar flares are
$< 10^{-7}$, $10^{-7}$-$10^{-6}$, $10^{-6}$-$10^{-5}$, $10^{-5}$-$10^{-4}$ and $\geq 10^{-4}$ W/m respectively.
Each X-ray class is divided into a linear scale from 1 to 9 (e.g. B1 to B9, C1 to C9, and M1 to M9).
An X17 class flare is 17 times more energetic than an X1 class flare.}

We removed long-term brightness variations around the flare peak
by fitting a quadratic function. We used the data from
0.05 to 0.01 (in the case of $n=3$) or from 0.15 to 0.03 ($n=10$)
days before the flare peak and the data
from 0.05 to 0.10 day ($n=3$) or 0.15 to 0.25 day ($n=10$) after 
the peak for the fitting.
After removing the long-term light variations, the flare start time
and flare end time are defined to be the time at which the residuals
of stellar brightness become larger and smaller than three times of 
the photometric error ($\sigma _{\rm photo}$; PDC flux errors given in the Kepler data) respectively.
We analyzed only flares with the duration longer than 5 min (at least 
5 consecutive data points exceed $3 \sigma _{\rm photo}$)
since it is difficult to distinguish between true flares and false events 
in the case of a single data point above the threshold (e.g. spikes in figure 1 (h)).

We excluded
the flare candidates which have longer rise time than decay time.
Since the number of stars observed in short-cadence mode is only $\sim 1500$,
the possibility that two or more stars exhibit flares at the same time is negligible.
Thus we also removed the pairs of flare events on different stars which have the same flare peak time.
After removing these false events, we checked all light curves of 
flare candidates by eye
and removed non-flare events, since our flare detection method misidentified
some eclipses or light curve jumps as flares.
We also examined the pixel level data of
flare stars in order to exclude the contamination of flares from nearby 
stars or other false events. If the spatial pixel-distributions of the brightness of a star
on CCDs are different between during the flare and at the quiescence,
the candidate of a flare is revealed to be a light variation of nearby 
source or a non-astrophysical event \citep[e.g.][]{Maehara2012}. We removed 5 events in this step.

We checked the Kepler Eclipsing Binary Catalog \citep{Prsa2011, Slawson2011, Kirk2013}  
and removed all flare candidates which occurred on the eclipsing binary (14 stars).
We also removed all flare candidates on the stars which are listed as
visual or spectroscopic binaries in \cite{Notsu2015a} (2 stars).
In total, we eliminated 152 flare candidates on these 16 stars.
Since the fraction of known binary systems would be small, there may be 
more unknown binary systems in the sample.
According to \cite{Notsu2015a}, the lower limit of binary fraction among superflare stars
based on high-dispersion spectroscopy is about 30 \%, 
which is not so different from the binary fraction of nearby solar-type stars 
\citep[e.g.][]{Duquennoy1991}.

Bolometric energy emitted by each flare was estimated 
from the stellar luminosity, flare amplitude and flare duration
with the same manner as \cite{Shibayama2013}.
The luminosity of each flare
star were estimated from the effective temperature and the stellar 
radius taken from \cite{Huber2014}.
The error in bolometric energy was estimated 
from errors in the temperature and stellar radius from \cite{Huber2014}.
Since the error in the temperature  is $\sigma _{T_{\rm star}} \sim 200$ K ($\sigma _{T_{\rm star}}/T_{\rm star}\sim 3$-$4$\%) and that in
the stellar radius is $\sigma _{R_{\rm star}}/R_{\rm star} \sim 10$ \%,  the typical error in bolometric energy
 would be $\sigma _{E_{\rm flare}}/E_{\rm flare} \sim 25$ \%.

We estimated periods of long-term light variations from light curves 
of each star by using the discrete Fourier transform (DFT) method.
For period analysis, we used the long-cadence data (time resolution of 30 min)
between 2009 September (Q2) and 2013 April (Q16).
After removing linear trends in each quarter and 
the correcting the difference in mean flux between different quarters,
we calculated the power spectrum and
chose a peak corresponding to the maximum power among peaks which are 20 times 
larger than the noise level.

\section{Results and discussion}
We detected 187 flares on 23 solar-type stars from the data of 1547
solar-type stars.
The details of all detected flares and properties of superflare stars are summarized in table 1-6 in Additional file 1 
(CSV version of table 1 and table 2-6 are also available as Additional file 2 and 3 respectively).
Figure 1 shows typical light curves of superflares from short-cadence data.
Light curves of all detected flares are shown in figure 1-94 in Additional file 1.
The amplitudes of flares in a unit of average brightness of the star
range from $1.3\times 10^{-3}$ to $8.5\times 10^{-2}$
and the bolometric 
energies of flares range from $2\times 10^{32}$ to $8\times 10^{35}$ erg. The lower end of 
the amplitude of detected flare is comparable to that of the X17 class 
solar flare (normalized amplitude $\sim 3 \times 10^{-4}$) occurred on Oct 28, 2003 \citep{Kopp2005}.
As shown in figure 1 (b) and (d), some flares show multiple peaks with 
the peak separation of 100 - 1000 seconds.
The time scale of these time variations is comparable to 
that of flare oscillations observed in 
the Sun \citep[e.g.][]{Aschwanden1999, Nakariakov2009} and 
M dwarfs 
\citep{Mitra2005, Welsh2006,Anfinogentov2013}.
On the analogy of the solar flare oscillations, 
these multiple peaks may be caused by coronal loop oscillations associated with flares.

\subsection{Occurrence frequency distribution}
Figure 2 represents the occurrence frequency distribution of superflares with energy $>10^{33}$ erg as a function of the rotation period.
The flare frequency strongly depends on the rotation period
and the frequency rapidly decreases as the rotation period 
increases in the period above a few days.
No flare was detected on the stars with $P_{\rm rot}\geq 20$ day from short-cadence data and 
the upper limit of the frequency for this bin is $<$ 1 in 80 years.
The flare occurrence frequency on the stars with $2.5 \leq P_{\rm rot} < 20$ day
roughly proportional to $P_{\rm rot}^{-3}$. 
The flare frequency on the stars with $P_{\rm rot}\geq 20$ day
expected from this decrease trend
 is $\sim 0.004$ year$^{-1}$ star $^{-1}$
 which is about 8 times lower than that
on the stars with $10 \leq P_{\rm rot} < 20$ day.
The similar relation between flare frequency and rotation period 
was reported by previous studies using the long-cadence data 
\citep[e.g.][]{Notsu2013b}.
The rotation period is thought to be related to the age of stars \citep[e.g.][]{Ayres1997}.
This result suggests that the young solar-type stars would exhibit superflares more frequently.

Figure 3 (a) represent the occurrence frequency of superflares as a function of the bolometric energy of superflares.
Solid histogram in figure 3 (a) represents the frequency distribution of superflares on all solar-type stars derived 
from short-cadence data and dashed histogram represents that from long-cadence data \citep{Shibayama2013}.
The average flare frequencies in a given energy bin from short-cadence data are about 5 times higher than those from long-cadence data.
This difference in the flare frequency is mainly caused by that in the rotation period distribution of the observed targets.
The period distribution of the targets in short-cadence mode is biased to the shorter-period end. 
The fraction of stars with the rotation period ($P_{\rm rot}$) $< 10$ days observed in long-cadence mode is 8.1 \% (8312/102709). 
On the other hands, that of stars with $P_{\rm rot} < 10$ days in short-cadence mode is 32\% (499/1547).
As mentioned above, the stars with shorter rotation period exhibit more frequent superflares.
Therefore, the average flare frequency derived from short-cadence data is 
higher than
that from long-cadence data.
Solid histogram in figure 3 (b) represents the corrected frequency distribution of superflares from 
short-cadence data taking into account the difference in the period distribution of the observed samples.
Since the ratio of total observation time (sum of the length of observation time for each star)
for stars with $P_{\rm rot}<10$ day to that for stars with $P_{\rm rot} \geq 10$ day
in short-cadence mode (1:1.7) is about 6.5 times larger than the ratio in long-cadence mode (1:11), 
we calculated the corrected occurrence frequency by adjusting the ratio of total observation time 
in short-cadence mode to that in long-cadence mode.
Both frequency distributions from short- and long-cadence data are almost the same for the flare energy between $10^{34}$ erg and $10^{36}$ erg.
The frequency of flares derived from the long-cadence data is less than that from the short-cadence data for the flare energy below $10^{34}$ erg.
This difference is caused by the detection limit of flare search method. Since the time-resolution of long-cadence data is much lower than that of short-cadence data, we could not detect smaller flares from long-cadence data and observed flare frequency from long-cadence data becomes much lower than that from short-cadence data.
The frequency distribution of superflares from the short-cadence data can be fitted by a power-law function with the index of $-1.4\pm 0.2$ for the energy between $4\times 10^{33}$ erg and $1\times 10^{36}$ erg. 
This value is comparable to the power-law index for the frequency distribution of superflares derived from the long-cadence data.
Using the combined data set from both short- and long-cadence data, the power-law index is $-1.5 \pm 0.1$
for the flare energy of $4\times 10^{33}$ - $1\times 10^{36}$ erg.

\cite{Shibata2013} pointed out that the frequency distribution of superflares on Sun-like stars (early-G dwarfs with $P_{\rm rot} \geq 10$ day)
and those of solar flares are roughly on the same power-law line.
Figure 4 represents the comparison between the frequency distribution of 
superflares on Sun-like stars ($5800 \leq T_{\rm eff} < 6300$ and $P_{\rm rot} \geq 10$ day) derived from both 
short- (filled circles) and long-cadence (solid line) data and those of solar flares (dashed lines).
The thin dotted line indicates the power-law function with an index of -1.8 taken from \cite{Shibata2013}.
The frequency distribution of superflares derived from short-cadence data is also on the same power-law line.
This distribution suggests that average occurrence frequency of superflares on Sun-like stars with bolometric energy of 
$10^{33}$ erg,  $10^{34}$ erg, and $10^{35}$ erg are once in $\sim 70$ years, $\sim 500$ years and $\sim 4000$ years respectively.
As mentioned in previous paragraph, the frequency of superflares strongly depends on the rotation period. The flare frequency on the
stars with the same rotation period as the Sun ($P_{\rm rot} \sim  25$ day)
would be $\sim 8$ times smaller that that on the stars with $10 \leq P_{\rm rot} < 20$ day.
Therefore the occurrence frequency of superflares with energy $> 10^{33}$ erg would be once in 500-600 years.
These values are lower than the upper limits derived from radionuclides in lunar samples \citep{Schrijver2012}.
\cite{Miyake2012} reported a rapid increase in the $^{14}$C content of tree rings around A.D. 774-775 and \cite{Miyake2013} found another $^{14}$C increase event around A.D. 993-994. The cosmic-ray flux of these events are equivalent to solar flares with the energy of $\sim 10^{34}$ ergs, if cosmic-ray sources of these events were solar flares. The occurrence frequency of these events (two events in 270 years) is roughly comparable to the average occurrence frequency of superflares on Sun-like stars with the energy of $10^{33}$ - $10^{34}$ erg.
In the case of solar flares, the fraction of geo-effective CMEs is less than $10$\% \citep[e.g.][]{Gopalswamy2007}.
If the fraction is same for superflares on our Sun, the occurrence rate of extreme GIC events (10 times or more larger than the Carrington event)
would be less than once in a few thousand years.

\subsection*{Relation between flare energy and area of starspots}
Most of superflare stars show large-amplitude light variations with periods
from a few days to $\sim 20$ days. 
If we assume these quasi-periodic variations are caused by the rotation of the star, the amplitude and period of light variations correspond to the area of starspots and the rotation period \citep{Notsu2013b}. 
The large-amplitude light variations of superflare stars suggest that the surface of superflare stars are corved by large starspots.

Figure 5 shows the scatter plot of flare energy as a function of the area of starspot group ($A_{\rm spot}$).
The area of starspots on superflare stars were estimated from the amplitude of light variations ($\Delta F_{\rm rot}$) by using the following equation:
\begin{linenomath}
\begin{equation}
A_{\rm spot} = \Delta F_{\rm rot} A_{\rm star} \left\{1-\left(\frac{T_{\rm spot}}{T_{\rm star}}\right)^{4}\right\}^{-1}, 
\end{equation}
\end{linenomath}
 where $A_{\rm star}$ is the apparent area of the star, $T_{\rm spot}$ and $T_{\rm star}$ are the temperature of starspot and photosphere of the star \citep{Notsu2013b}.
Here we define the
amplitude as the brightness range normalized by the average stellar brightness, in which the lower 99\% of the
distribution of the brightness difference from average, except for the flares, are
included. We estimated the amplitude from the long-cadence data of the quarter at which each superflares detected from the short-cadence data occurred.
$A_{\rm star}$ and $T_{\rm star}$ for each superflare star were taken from \cite{Huber2014} and we assumed $T_{\rm spot} = 4000$K \citep{Berdyugina2005}.
Although there is large scatter, the majority of energetic flares occur on the stars with large starspots.
Since the flares are a result of a sudden conversion of magnetic energy to other forms \cite[i.e. thermal and kinetic energy; e.g.][]{Shibata2011},
the flare energy is supplied from the magnetic energy stored near the starspots ($E_{\rm mag}$) 
and
the bolometric energy released by the flare may be a fraction of the magnetic energy
\citep{Maehara2012, Shibata2013, Notsu2013b}.
Here the stored magnetic energy can be roughly estimated by
\begin{linenomath}
\begin{equation}
E_{\rm mag} \sim \frac{B^2 L^3}{8\pi} ,
\end{equation}
\end{linenomath}
where $B$ and $L$ correspond to the magnetic field strength of the starspots 
and the scale length of the starspot group respectively.
By using the 
total area of starspot group  $A_{\rm spot}\sim L^{2}$, the energy released by a flare can be written as
\begin{linenomath}
\begin{equation}
E_{\rm flare} \sim f E_{\rm mag} \sim \frac{f B^2 A_{\rm spot}^{3/2}}{8\pi} ,
\end{equation}
\end{linenomath}
where $f$ is the fraction of magnetic energy released by the flare.
This suggests that the flare energy is roughly proportional to the area of
the starspot group to the power of 3/2.
Since $f=0.1$ \citep{Aschwanden2014}, 
the typical sunspot area for generating X10-class flares ($\sim 10^{32}$ erg)
observed in 1989-1997  was $3\times  10^{-4}$ of the half area of the solar surface \citep{Sammis2000}
and the typical magnetic field strength of sunspot is the order of $1000$ G \citep[e.g.][]{Antia2003}, 
equation (5) can be written as
\begin{linenomath}
\begin{equation}
E_{\rm flare} \sim 7\times 10^{32} {\rm (erg)} \left(\frac{f}{0.1}\right) \left(\frac{B}{1000{\rm G}}\right)^2 \left[\frac{A_{\rm spot}/2\pi R_{\odot}^2}{0.001}\right]^{3/2} .
\end{equation}
\end{linenomath}
Solid and dashed lines in the figure 5 represent equation (6) for $f=0.1$, $B=1000$ and $3000$ G respectively. 
Majority of superflares detected from short-cadence data (filled-squares) and almost all solar flares (small dots) are below these lines. 
This suggests that the upper limit of the energy released by the flare is basically determined 
by the magnetic energy estimated from the area of starspots.
However the bolometric energies of superflares detected from long-cadence data (small-crosses) are biased to
larger-energy end. As mentioned above, due to the low time-resolution,
only superflares with larger bolometric energies could be detected from long-cadence data. In addition, superflares 
with large energy but small occurrence frequency 
could be detected from the long-cadence data because of the 
large number of stars ($\sim 10^5$).

There are some superflares above the analytic relation.
KIC 7093428 is one of superflare stars exhibiting superflares whose energy is much larger than that expected from equation (6).
This object is located within the error circle of an X-ray source 1SXPS J184925.7+423901 \citep{Evans2014}.
The X-ray luminosity of the object is $\sim 10^{28}$ erg/s if the distance of the object is $\sim 200$ pc. 
The empirical relation between the X-ray luminosity and the magnetic flux of the star \citep{Pevtsov2003} 
suggests that the magnetic flux of KIC 7093428 is the order of $10^{24}$ Mx.
This value corresponds to the spot area of $3\times 10^{20}$ cm$^{2}$ for $B=3000$G,
and the energy of the largest superflare on KIC 7093428 is 
comparable to the flare energy estimated from equation (6).
The spot area of the object expected from the X-ray luminosity is more than one order of magnitude larger than that estimated from the amplitude of quasi-periodic modulations.
The amplitude of light variations due to the rotation of star is affected by the inclination of rotation axis and the latitude of starspots \citep[e.g.][]{Notsu2013a, Notsu2013b}.
In the case of stars with low-inclination angle or stars with starspots around polar region, the light variation caused by the rotation become small
and the area of starspots derived from the amplitude of light variation is smaller than the actual area.
According to \cite{Notsu2015b}, some superflare stars which show superflares with energy
larger than that expected from the amplitude of light variation have low-inclination angles.

In addition, 
polar spots are often observed on rapidly-rotating cool stars \citep[e.g.][]{Strassmeier2009}.
Another possibility is flares on the companion star.
\cite{Kitze2014} found that the astrometric position of the Sun-like superflare star, KIC 7133671
which show a superflare with energy larger than expected from the photometric spot area, was shifted by 25 mas during the flare. 
This suggests that the flare on KIC 7133671 occurred on a faint companion 
star. In this case, the flare energy would not be related to the area of 
starspot group on the primary star.

\subsection{Correlation between flare duration and flare energy}
Figure 6 shows the duration of superflares as a function of the bolometric
energy of flares.
Here we defined the flare duration as the $e$-folding decay time of flare intensity after its peak.
The duration of superflares increases as the flare energy increases.
The linear fit for the superflares from short-cadence data in the 
$\log$-$\log$ plot yields
\begin{linenomath}
\begin{equation}
\tau _{\rm flare} \propto E^{0.39\pm 0.03}_{\rm flare}
\end{equation}
\end{linenomath}
where $\tau _{\rm flare}$ and $E_{\rm flare}$ indicate the duration and bolometric energy of flares.
The duration of superflares derived from long-cadence data (crosses) is longer than that from short-cadence data. 
This difference may be caused by the difference in time resolution of the data and selection bias. 
Since the time resolution of long-cadence data is 30 min and much longer than 
that of short cadence data (1 min), the flare detection method for long-cadence data used in \cite{Maehara2012} and \cite{Shibayama2013}
can detect only superflares with long duration.
The bolometric energy and duration in optical wavelength of the X17
solar flare on 2003 October 28 are 
the order of $10^{32}$ erg and $\sim 10$ min \citep{Kopp2005}, which are
roughly comparable to the correlation between the duration and energy of superflares on solar-type stars.

Similar correlation between the flare duration and energy was observed in solar flares.
The power-law slope for the correlation between the duration of solar flares and X-ray fluence observed with the GOES is about 1/3 \citep{Veronig2002}.
The correlation between the duration and peak flux observed with RESSI also shows similar power-law slope of $\sim 0.2$ \citep{Christe2008}.
These similarity between solar flares and superflares on solar-type stars suggests that solar flares and superflares are 
caused by the same physical process (i.e. reconnection).

As discussed in the previous subsection, the flare energy is related to the magnetic energy stored near the starspots as follows:
\begin{linenomath}
\begin{equation}
E_{\rm flare} \sim f E_{\rm mag} \propto f B^{2} L^{3},
\end{equation}
\end{linenomath}
where $f$, $B$, and $L$ correspond to the fraction of energy released by a flare, 
the magnetic field strength of the starspots
and the scale length of the starspot group respectively.
On the other hand, the duration of flares, especially the duration of impulsive phase of flares,
is comparable to the reconnection time ($\tau _{\rm rec}$) which can be written as
\begin{linenomath}
\begin{equation}
\tau _{\rm flare} \sim \tau _{\rm rec} \sim \tau _{\rm A}/M_{\rm A} \sim L/v_{\rm A}/M_{\rm A},
\end{equation}
\end{linenomath}
where $\tau _{\rm A} = L/v_{\rm A}$ is the Alfv$\acute{\rm e}$n time, 
$v_{\rm A}$ is the Alfv$\acute{\rm e}$n velocity, and $M_{\rm A}$ is
the non-dimensional reconnection rate which is comparable to 0.1-0.01 for the fast
reconnection such as Petschek model \citep[e.g.][]{Shibata2011}.
Since all G-type main sequence stars have similar stellar properties (e.g. $T_{\rm eff}$ and $\log g$), 
$B$ and $v_{\rm A}$ might not be so different among them.
Therefore, from equation (8) and (9), the duration of flares can be written as
\begin{linenomath}
\begin{equation}
\tau _{\rm flare} \propto E_{\rm flare} ^{1/3}.
\end{equation}
\end{linenomath}
This suggests that the power-law slope for the correlation between the flare duration 
and flare energy is about $1/3$ and this is comparable to the observed value of $0.39\pm 0.03$.

\section{Conclusions}
We found 187 superflares on 23 solar-type stars from the 1-min cadence data 
obtained with the Kepler space telescope. Because of the high temporal resolution of
the data, we found the following new results.
\begin{enumerate}
\item The power-law frequency distribution of superflares \citep[e.g.][]{Maehara2012, Shibayama2013}
continues to the flare energy of $10^{33}$ erg and the average occurrence rate of the X100 
superflares on the stars with similar rotation period to the Sun is about once in 500-600 years.
\item Some of superflares show multiple peaks during a flare with the peak separation of $100$-$1000$ seconds.
The time scale of these modulation during the flare is comparable to the periods of quasi-periodic pulsations in
solar and stellar flares.
\item The flare duration increases with the flare energy as $\tau_{\rm flare} \propto E_{\rm flare}^{0.39\pm 0.03}$.
\end{enumerate}
Observations of solar flares, especially white light flares, with
the high temporal resolution are also important to compare the properties of solar flares
and those of superflares on solar-type stars.

%
%
\section{Competing interests}
The authors have no competing interests to declare.

\section{Authors' contributions}
H.M. developed the flare detection method and was responsible for data analysis. 
He also wrote the majority of the text. 
T.S. contributed to the analysis of stellar flares observed by the Kepler.
S.N. and Y.N. analyzed the rotation of stars observed by the Kepler.
S.H. and D.N. contributed to the interpretation of the stellar brightness variation 
and helped the data analysis from general stellar astronomical point of view. 
K.S. provided theoretical interpretation of the observations and provided advice on the paper's content.


\acknowledgments{%
Kepler was selected as the tenth Discovery mission. 
Funding for this mission is provided by the NASA Science Mission Directorate. 
The data presented in this paper were obtained from the Multimission Archive at STScI. 
This work was supported by MEXT/JSPS KAKENHI Grant Number 25287039, 26400231, and 26800096.
}

{}

\noindent
\textbf{Figure legends}\\

\includegraphics[width=0.9\hsize]{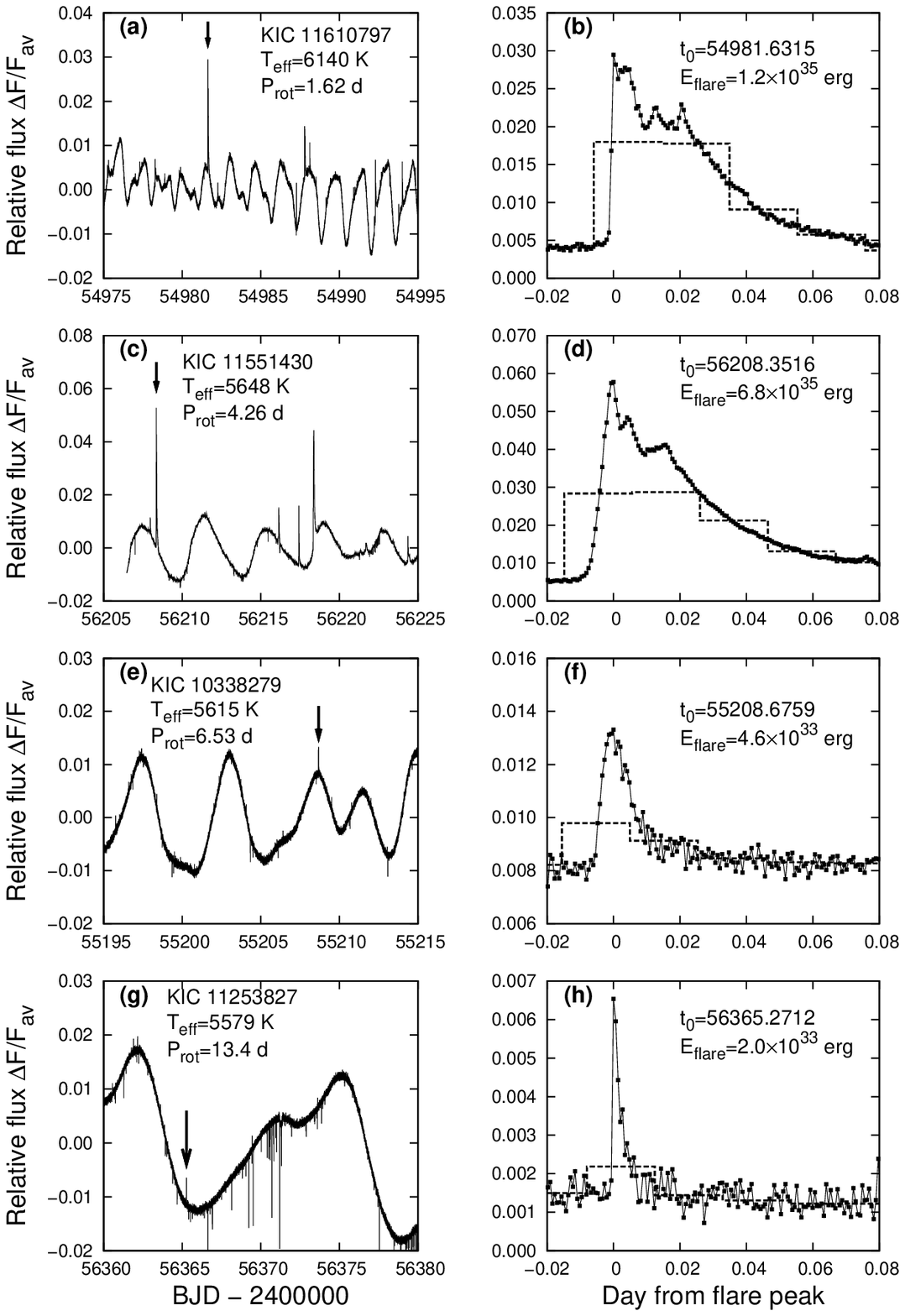}\\
\textbf{Figure 1. 
Light curves of typical superflares.
}
\textbf{(a)} Long-term light variations of the G-dwarf KIC 11610797. The vertical axis
means the relative difference between observed brightness of the star
and the average brightnes during the observation period.
The horizontal axis means the times of the observations in 
Barycentric Julian Date.
\textbf{(b)} Enlarged light curve of a superflare observed around Barycentric Julian Date (BJD) 2454981.63
(indicated by the down arrow in panel (a)). 
Filled-squares with solid-lines and dashed-lines represent the light curves from short- and long-cadence data respectively. The long-cadence light curve is shifted in vertical direction for clarity.

The flare peak ($t_{0}$) is BJD 2454981.6315 and the bolometric
energy of this flare ($E_{\rm flare}$) is $1.2\times 10^{35}$ erg.
\textbf{(c)} Same as panel (a), but for KIC 11551430.
\textbf{(d)} Same as panel (b), but for the superflare observed on KIC 11551430 around
BJD 2456208.35 (down arrow in panel (c)).
\textbf{(e)} Same as panel (a), but for KIC 10338279.
\textbf{(f)} Same as panel (b), but for the superflare observed on KIC 4543412 around
BJD 2455208.68 (down arrow in panel (e)).
\textbf{(g)} Same as panel (a), but for KIC 11253827.
\textbf{(h)} Same as panel (b), but for the superflare observed on KIC 11253827 around
BJD 2456365.27 (down arrow in panel (g)).

\includegraphics[width=0.9\hsize]{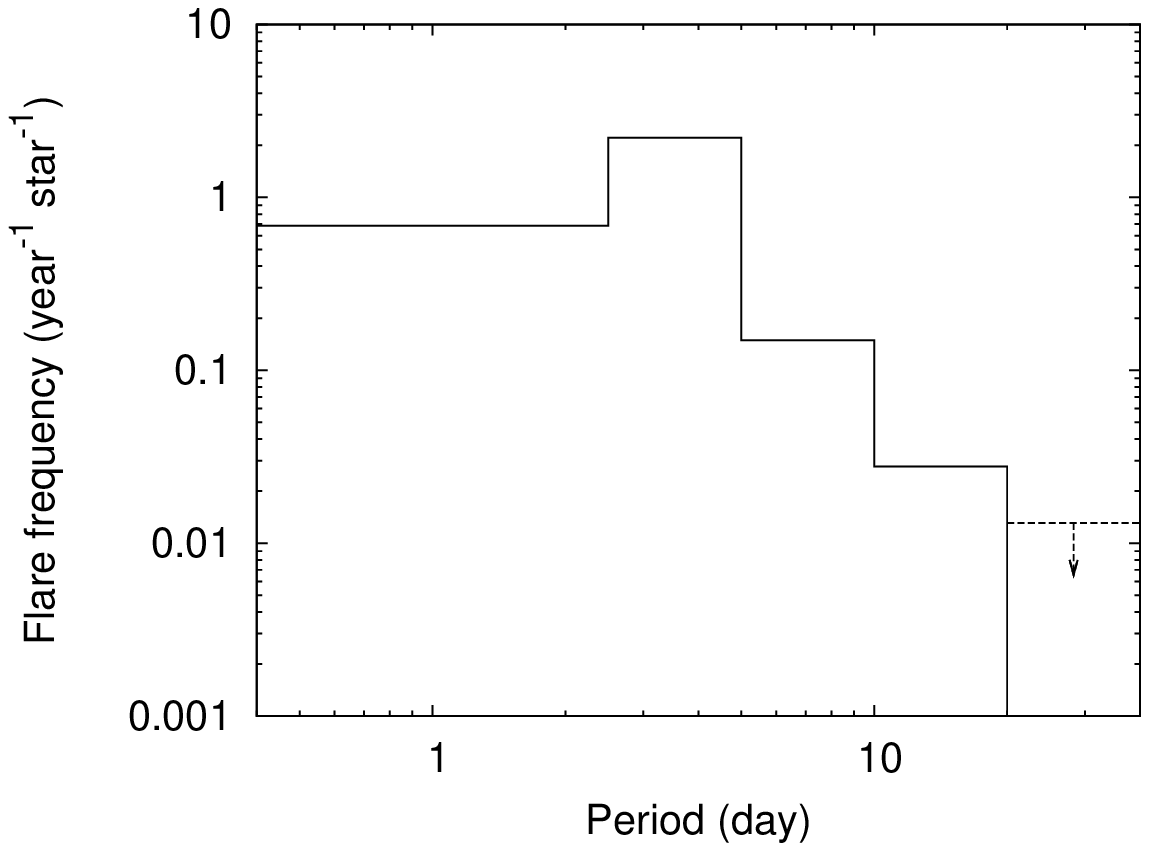}\\
\textbf{Figure 2. Occurrence frequency distribution of superflares as a function of the rotation period.}
The horizontal axis indicates the rotation period of the star derived from photometric data. The vertical axis indicates the number of superflares per star and per year in each period bin.
Period bins are defined as follows: $P < 2.5$ day , $2.5\leq P < 5$ day, $5\leq P < 10$ day, $10 \leq P < 20$ day, and $P \geq 20$ day.
Solid-line histogram represents the occurrence frequency of superflares with an energy of $>10^{33}$ erg in each period bin. The dashed down-arrow indicates the upper limit of the flare frequency.

\includegraphics[width=0.9\hsize]{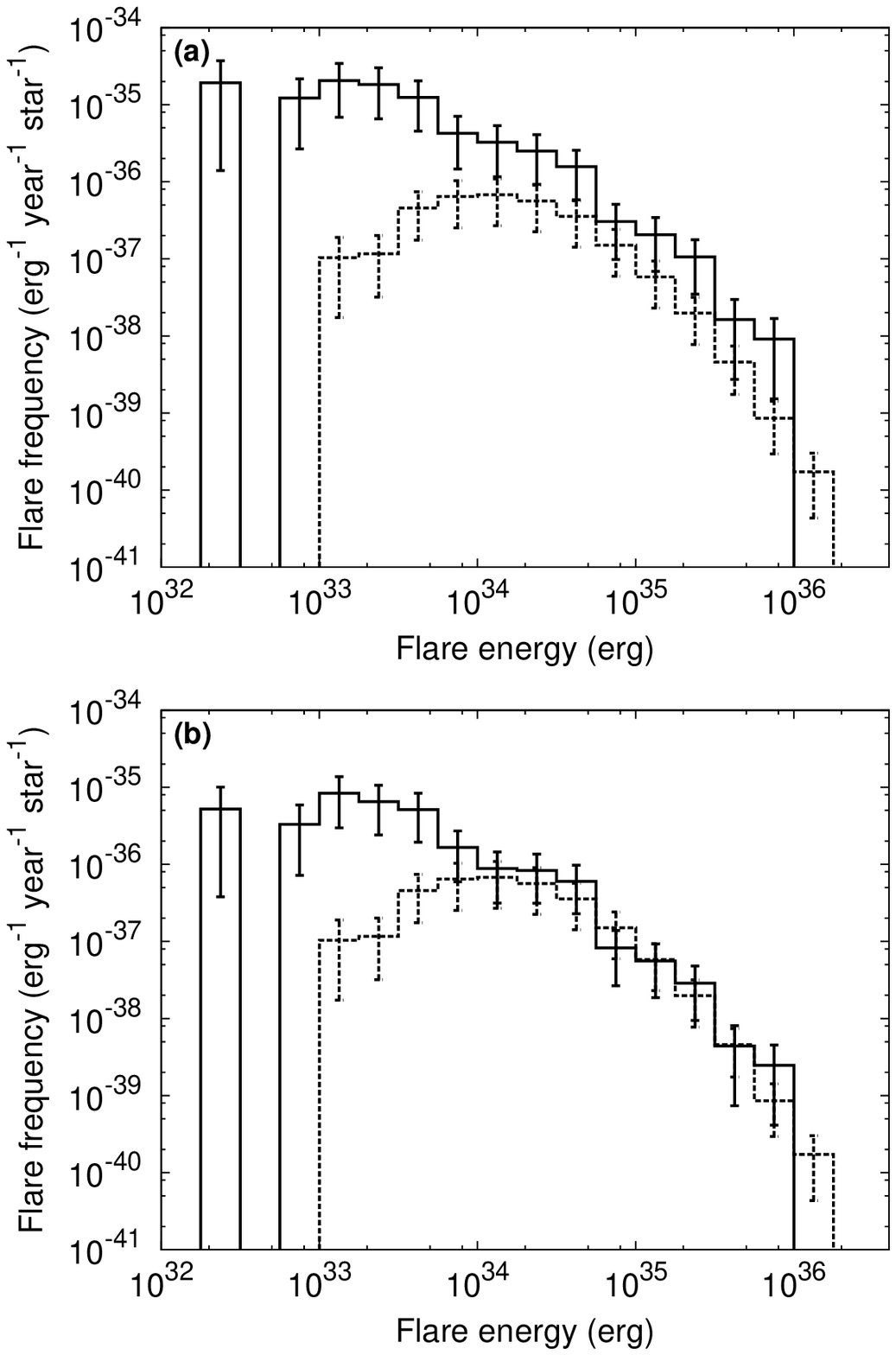}\\
\textbf{Figure 3. Occurrence frequency distribution of superflares as a function of flare energy.}
(a)
Bold solid and dashed histograms represent the occurrence frequency of superflares on all G-type main sequence stars 
detected from short- (this work) and long-cadence data \citep{Shibayama2013}as a function of total energy of superflares respectively.
The vertical axis indicates the number of superflares per star, per year, and per unit energy.
Error bars represent 1-$\sigma$ uncertainty in the frequency estimated from the square-root of the number of detected flares in each energy bin and the uncertainty in flare energy.
(b)
Same as (a), but for the corrected frequency distribution from the short-cadence data.
In long-cadence mode, the ratio of total observation time (number of stars $\times $ observation time of each star) 
for stars with $P_{\rm rot}<10$ day to that for stars with $P_{\rm rot} \geq 10$ day is about 1:11.
On the other hand, the ratio in short-cadence mode is about 1:1.7.
The corrected occurrence frequency was calculated so that 
the ratio of total observation time in short-cadence mode is 
same as that in long-cadence mode.
\newpage

\includegraphics[width=0.9\hsize]{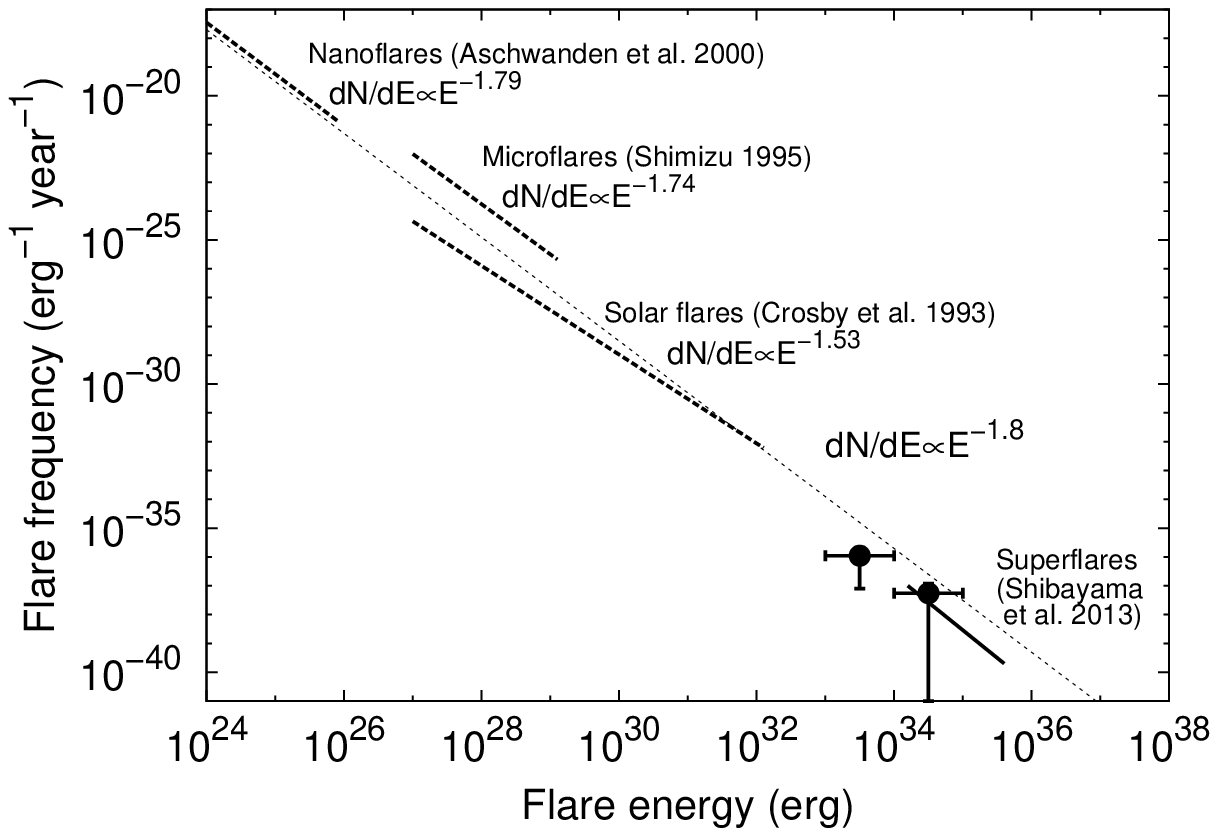}\\
\textbf{Figure 4. Comparison between occurrence frequency superflares on Sun-like stars and those of solar flares.}
Filled-circles indicates the occurrence frequency distribution of superflares on Sun-like stars (G-type main sequence stars
with $P_{\rm rot}>10$ days and $5800<T_{\rm eff}<6300$K) derived from short-cadence data.
Horizontal error bars represent the range of each energy bin.
The definition of the vertical error bars is the same as figure 3.
Bold-solid line represents the power-law frequency distribution of superflares on Sun-like 
stars taken from \cite{Shibayama2013}.
Dashed lines indicate the power-law frequency distribution of solar flares observed in 
hard X-ray \citep{Crosby1993}, soft X-ray \citep{Shimizu1995}, and EUV \citep{Aschwanden2000}.
Occurrence frequency distributions of superflares on Sun-like stars and solar flares are roughly on the same power-law line 
with an index of $-1.8$ (thin-solid line) for the wide energy range between $10^{24}$ erg and $10^{35}$ erg.

\includegraphics[width=0.9\hsize]{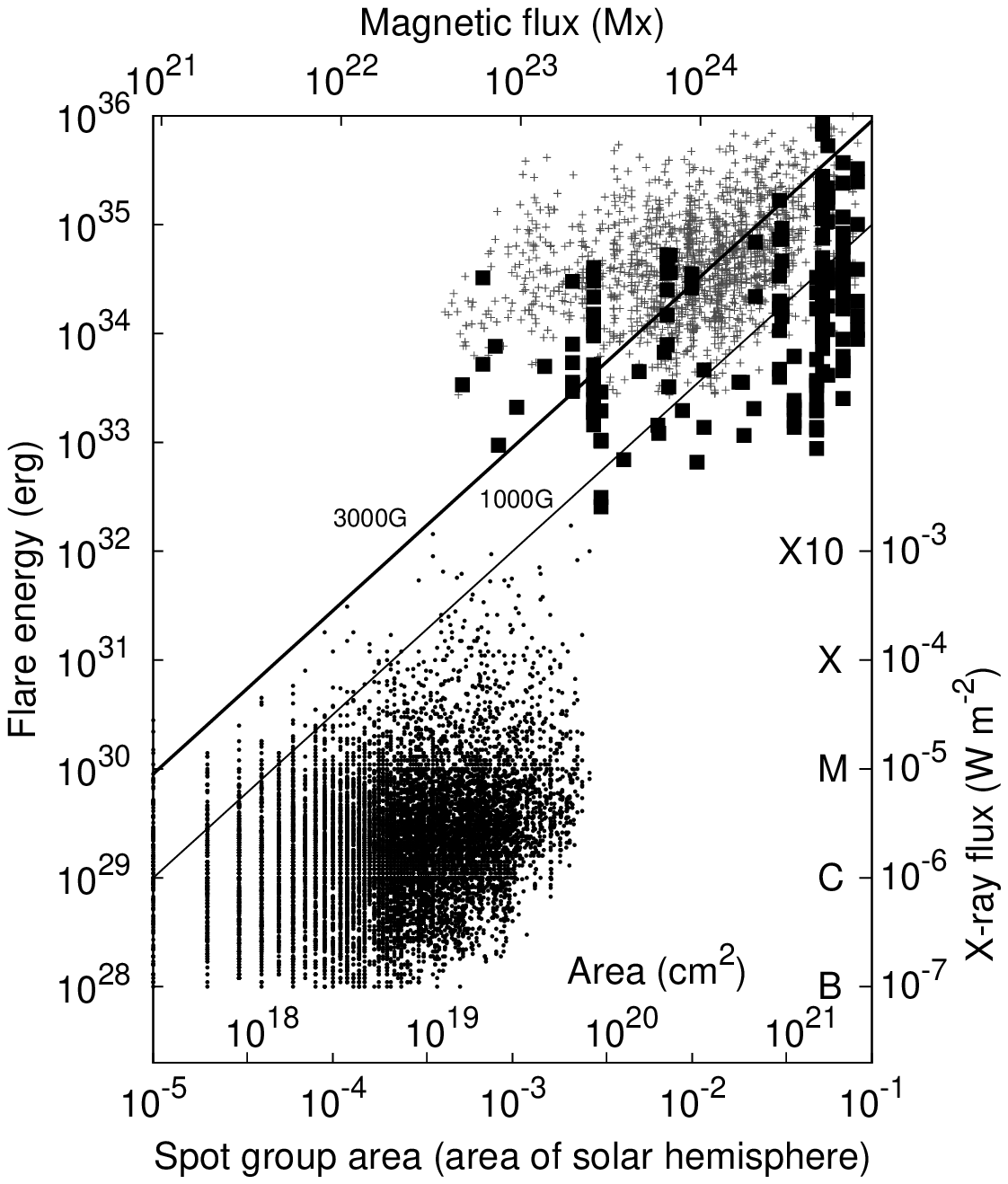}\\
\textbf{Figure 5. Scatter plot of flare energy as a function of spot area.}
The lower and upper horizontal axis indicate the area of starspot group in the unit of the area of solar hemisphere and the magnetic flux for $B=3000$ G.
The vertical axis represents the bolometric energy released by each flare.
Filled-squares and  small-crosses indicate 
superflares on G-type main sequence stars
detected from short-(this work) and long-cadence data \citep{Shibayama2013} respectively.
Small filled-circles 
represent solar flares 
(Ishii et al., private communication, based on the data
retrieved from the website of the National Geophysical 
Data Center of the National Oceanic and Atmospheric Administration 
(NOAA/NGDC), Solar-Terrestrial Physics Division at
http://www.ngdc.noaa.gov/stp/).
The area of starspot group for G-type main sequence stars was estimated from the full amplitude of light variations
during each observation period in which the flare occurred
 using equation (3).
We assumed that bolometric energies of B, C, M, X, and X10 class solar flares 
are $10^{28}$, $10^{29}$, $10^{30}$, $10^{31}$, and $10^{32}$ erg
from observational estimates of energies of typical solar flares
\citep[e.g.][]{Benz2008, Emslie2012}.

\includegraphics[width=0.9\hsize]{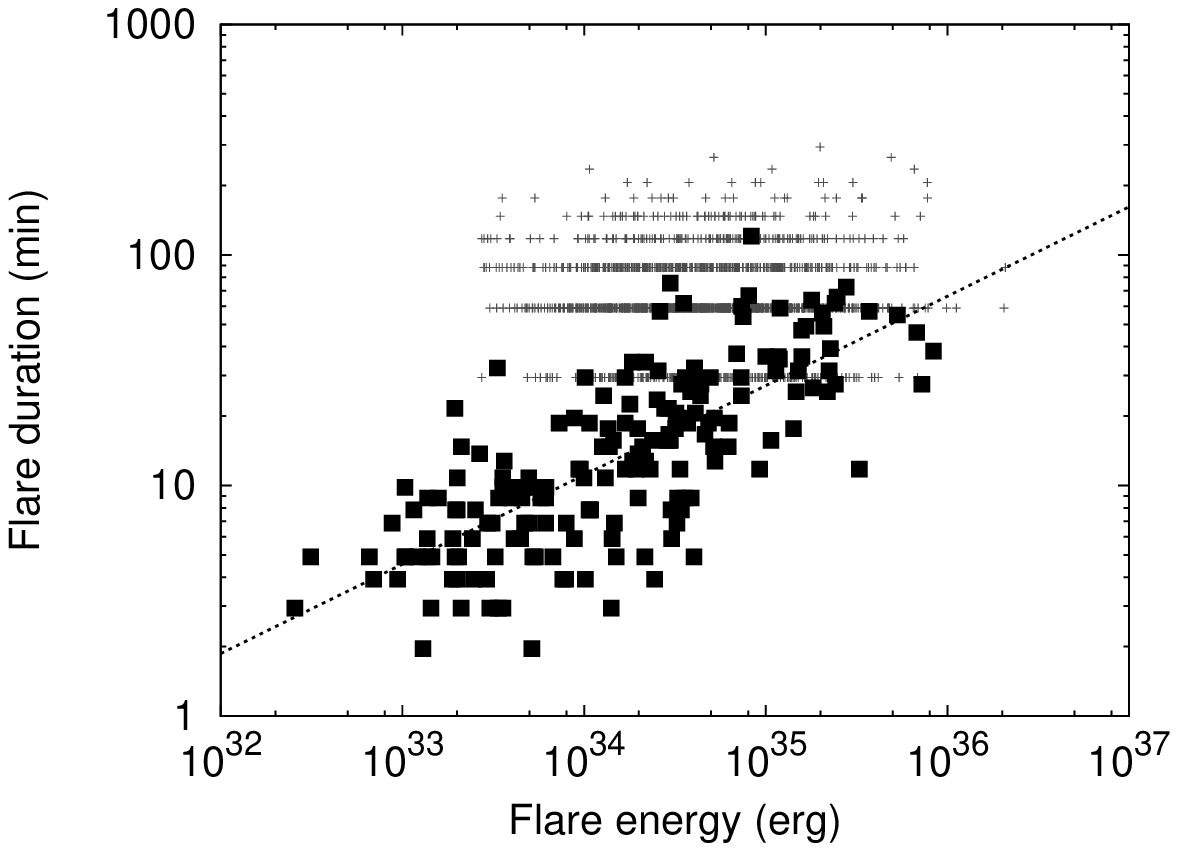}\\
\textbf{Figure 6. Scatter plot of the duration of superflares as a function of the bolometric energy.}
Filled-squares and small-crosses indicate superflares on G-type main sequence stars detected from short- and long-cadence data respectively.
We used $e$-folding decay time as the flare duration.
Dotted line indicates the linear regression for the data of superflares from short-cadence data. The power-law slope of the line is $0.39 \pm 0.03$.

\begin{thebibliography}{}
\bibitem[{{Allen et al.}(1989)}]{Allen1989}
Allen, J., Sauer, H., Frank, L., \& Reiff, P. (1989)
Effects of the March 1989 solar activity.
Eos, Transactions American Geophysical Union 70: 1479–1488. doi:10.1029/89EO00409

\bibitem[{{Anfinogentov et al.}(2013)}]{Anfinogentov2013}
Anfinogentov, S., 
Nakariakov, V.~M., Mathioudakis, M., Van Doorsselaere, T., 
\& Kowalski, A.~F. (2013)
The Decaying Long-period Oscillation of a Stellar Megaflare.
The Astrophysical Journal 773: 156. doi:10.1088/0004-637X/773/2/156

\bibitem[{{Antia et al.}(2003)}]{Antia2003}
Antia, H.~M., Bhatnagar, A., \& Ulmschneider, P. (2003)
Lectures on Solar Physics.
Springer Verlag, Berlin


\bibitem[{{Aschwanden et al.}(1999)}]{Aschwanden1999}
Aschwanden, M.~J., Fletcher, L., Schrijver, C.~J., \& Alexander, D. (1999)
Coronal Loop Oscillations Observed with the Transition Region and Coronal Explorer.
The Astrophysical Journal 520: 880-894. doi:10.1086/307502

\bibitem[{{Aschwanden et al.}(2000)}]{Aschwanden2000}
Aschwanden, M.~J., Tarbell, T.~D., Nightingale, R.~W., et al. (2000)
Time Variability of the ``Quiet'' Sun Observed with TRACE. II. Physical Parameters, Temperature Evolution, and Energetics of Extreme-Ultraviolet Nanoflares.
The Astrophysical Journal 535: 1047-1065. doi:10.1086/308867

\bibitem[{{Aschwanden et al.}(2014)}]{Aschwanden2014}
Aschwanden, M.~J., Xu, Y., \& Jing, J. (2014)
Global Energetics of Solar Flares. I. Magnetic Energies.
The Astrophysical Journal 797: 50. doi:10.1088/0004-637X/797/1/50

\bibitem[{{Ayres}(1997)}]{Ayres1997}
Ayres, T.~R. (1997)
Evolution of the solar ionizing flux.
Journal of Geophysical Research 102: 1641-1652. doi:10.1029/96JE03306

\bibitem[{{Benz}(2008)}]{Benz2008}
Benz, A.~O. (2008)
Flare Observations
Living Reviews in Solar Physics 5: 1. doi:10.12942/lrsp-2008-1


\bibitem[{{Berdyugina}(2005)}]{Berdyugina2005}
Berdyugina, S.~V.\ (2005)
Starspots: A Key to the Stellar Dynamo.
Living Reviews in Solar Physics 2: 8.  doi:10.12942/lrsp-2005-8

\bibitem[{{Brown et al.}(2011)}]{Brown2011}
Brown, T. M., Latham, D. W., Everett, M. E., Esquerdo, G. A. (2011)
Kepler Input Catalog: Photometric Calibration and Stellar Classification.
The Astrophysical Journal 142: 112. doi:10.1088/0004-6256/142/4/112


\bibitem[{{Candelaresi et al.}(2014)}]{Candelaresi2014}
Candelaresi, S., Hillier, A., Maehara, H., et al. (2014)
Superflare occurrence and energies on G, K and M type dwarfs.
The Astrophysical Journal 792: 67. doi:10.1088/0004-637X/792/1/67
 
\bibitem[{{Carrington}(1859)}]{Carrington1859}
Carrington, R. C. (1859)
Description of a Singular Appearance seen in the Sun on September 1, 1859.
Monthly Notices of the Royal Astronomical Society 20: 13-15

\bibitem[{{Christe et al.}(2008)}]{Christe2008}
Christe, S., Hannah, I.~G., Krucker, S., et al (2008)
RHESSI Microflare Statistics. I. Flare-Finding and Frequency Distributions.
The Astrophysical Journal 677: 1385-1394. doi:10.1086/529011

\bibitem[{{Crosby et al.}(1993)}]{Crosby1993}
Crosby, N.~B., Aschwanden, M.~J., Dennis, B.~R. (1993)
Frequency distributions and correlations of solar X-ray flare parameters.
Solar Physics 143: 275-299. doi:10.1007/BF00646488

\bibitem[{{Duquennoy \& Mayor}(1991)}]{Duquennoy1991}
Duquennoy, A., \& Mayor, M. (1991)
Multiplicity among solar-type stars in the solar neighbourhood. II - Distribution of the orbital elements in an unbiased sample.
Astronomy and Astrophysics 248: 485-524.

\bibitem[{{Emslie et al.}(2012)}]{Emslie2012}
Emslie, A.~G., Dennis, B.~R., Shih, A.~Y., et al. (2012)
Global Energetics of Thirty-eight Large Solar Eruptive Events.
The Astrophysical Journal 759: 71. doi:10.1088/0004-637X/759/1/71

\bibitem[{{Evans et al.}(2014)}]{Evans2014}
Evans, P.~A., Osborne, J.~P., Beardmore, A.~P., et al. (2014)
1SXPS: A Deep Swift X-Ray Telescope Point Source Catalog with Light Curves and Spectra.
The Astrophysical Journal Supplement 210: 8. doi:10.1088/0067-0049/210/1/8

\bibitem[{{Gilliland et al.}(2010)}]{Gilliland2010}
Gilliland, R.~L., Jenkins, J.~M., Borucki, W.~J., et al. (2010)
Initial Characteristics of Kepler Short Cadence Data.
The Astrophysical Journal Letters 713: L160-L163. doi:10.1088/2041-8205/713/2/L160

\bibitem[{{Gopalswamy et al.}(2007)}]{Gopalswamy2007}
Gopalswamy, N.,  Yashiro, S., \& Akiyama, S. (2007)
Geoeffectiveness of halo coronal mass ejections.
Journal of Geophysical Research (Space Physics) 112: 6112. doi:10.1029/2006JA012149

\bibitem[{{Huber et al.}(2014)}]{Huber2014}
Huber, D., Silva Aguirre, V., Matthews, J.~M., et al. (2014)
Revised Stellar Properties of Kepler Targets for the Quarter 1-16 Transit Detection Run.
The Astrophysical Journal Supplement 211: article id. 2. doi:10.1088/0067-0049/211/1/2

\bibitem[{{Koch et al.}(2010)}]{Koch2010}
Koch, D.~G., Borucki, W.~J., Basri, G., et al. (2010)
Kepler Mission Design, Realized Photometric Performance, and Early Science.
The Astrophysical Journal Letters 713: L79-L86. doi:10.1088/2041-8205/713/2/L79

\bibitem[{{Kopp et al.}(2005)}]{Kopp2005}
Kopp, G., Lawrence, G., Rottman, G. (2005)
The Total Irradiance Monitor (TIM): Science Results.
Solar Physics 230: 129-139. doi:10.1007/s11207-005-7433-9

\bibitem[{Kirk et al.}(2013)]{Kirk2013}
Kirk et al (2013); in prep
Kepler Eclipsing Binary Catalog - Third Revision (Beta) 
http://keplerebs.villanova.edu/. Accessed 16 Sep 2014

\bibitem[{{Kitze et al.}(2014)}]{Kitze2014}
Kitze, M., Neuh{\"a}user, R., Hambaryan, V., \& Ginski, C. (2014)
Superflares on the slowly rotating solar-type stars KIC10524994 and KIC07133671?
Monthly Notices of the Royal Astronomical Society 442: 3769-3776. doi:10.1093/mnras/stu1114

\bibitem[{{Landini et al.}(1986)}]{Landini1986}
Landini, M., Monsignori Fossi, B.~C., Pallavicini, R., \& Piro, L. (1986)
EXOSAT detection of an X-ray flare from the solar type star Pi-prime UMa
Astronomy and Astrophysics 157: 217-222

\bibitem[{{Loomis}(1861)}]{Loomis1861}
Loomis, E. (1861)
The great auroral exhibition of August 28th to September 4th 1859.
The American Journal of Science and Arts,  Second Series 32: 318-335

\bibitem[{{Maehara et al.}(2012)}]{Maehara2012}
Maehara, H., Shibayama, T., Notsu, S., et al. (2012)
Superflares on solar-type stars.
Nature 485: 478-481. doi:10.1038/nature11063

%
\bibitem[{{Mitra-Kraev et al.}(2005)}]{Mitra2005}
Mitra-Kraev, U., Harra, L.~K., Williams, D.~R., \& Kraev, E. (2005)
The first observed stellar X-ray flare oscillation: Constraints on the flare loop length and the magnetic field.
Astronomy and Astrophysics 436: 1041-1047. doi:10.1051/0004-6361:20052834

\bibitem[{{Miyake et al.}(2012)}]{Miyake2012}
Miyake, F., Nagaya, K., Masuda, K., \& Nakamura, T. (2012)
A signature of cosmic-ray increase in AD 774-775 from tree rings in Japan.
Nature 486: 240-242. doi: 10.1038/nature11123

\bibitem[{{Miyake et al.}(2013)}]{Miyake2013}
Miyake, F., Masuda, K., \& Nakamura, T. (2013)
Another rapid event in the carbon-14 content of tree rings.
Nature Communications 4: 1748. doi:10.1038/ncomms2783


\bibitem[{{Nakariakov \& Melnikov}(2009)}]{Nakariakov2009}
Nakariakov, V.~M., \& Melnikov, V.~F. (2009)
Quasi-Periodic Pulsations in Solar Flares.
Space Science Reviews 149: 119-151. doi:10.1007/s11214-009-9536-3

\bibitem[{{Nogami et al.}(2014)}]{Nogami2014}
Nogami, D., Notsu, Y., Honda, S., et al. (2014)
Two sun-like superflare stars rotating as slow as the Sun.
Publications of the Astronomical Society of Japan 66: L4. doi:10.1093/pasj/psu012

\bibitem[{{Notsu et al.}(2013a)}]{Notsu2013a}
Notsu, S., Honda, S., Notsu, Y., et al. (2013a)
High-Dispersion Spectroscopy of the Superflare Star KIC 6934317.
Publications of the Astronomical Society of Japan 65: 112. doi:10.1093/pasj/65.5.112

\bibitem[{{Notsu et al.}(2013b)}]{Notsu2013b}
Notsu, Y., Shibayama, T., Maehara, H., et al. (2013b)
Superflares on Solar-type Stars Observed with Kepler II. Photometric Variability of Superflare-generating Stars: A Signature of Stellar Rotation and Starspots.
The Astrophysical Journal 771: 127. doi:10.1088/0004-637X/771/2/127 

\bibitem[{{Notsu et al.}(2015a)}]{Notsu2015a}
Notsu, Y., Honda, S., Maehara, H., et al. (2015a)
High Dispersion Spectroscopy of Solar-type Superflare Stars. I. Temperature, Surface Gravity, Metallicity, and $v \sin i$. 
Publications of the Astronomical Society of Japan 67: in press. doi:10.1093/pasj/psv001

\bibitem[{{Notsu et al.}(2015b)}]{Notsu2015b}
Notsu, Y., Honda, S., Maehara, H., et al. (2015b)
High Dispersion Spectroscopy of Solar-type Superflare Stars. II. Stellar Rotation, Starspots, and Chromospheric Activities.
Publications of the Astronomical Society of Japan 67: in press. doi:10.1093/pasj/psv002

\bibitem[{{Pevtsov et al.}(2003)}]{Pevtsov2003}
Pevtsov, A.~A., Fisher, G.~H., Acton, L.~W., et al. (2003)
The Relationship Between X-Ray Radiance and Magnetic Flux.
The Astrophysical Journal 598: 1387-1391. doi:10.1086/378944

\bibitem[{Pr{\v s}a et al.}(2011)]{Prsa2011}
Pr{\v s}a, A., Batalha, N., Slawson, R.~W., et al. (2011)
Kepler Eclipsing Binary Stars. I. Catalog and Principal Characterization of 1879 Eclipsing Binaries in the First Data Release
The Astronomical Journal 141: 83. doi:10.1088/0004-6256/141/3/83

\bibitem[{{Sammis et al.}(2000)}]{Sammis2000}
Sammis, I., Tang, F., \& Zirin, H. (2000)
The Dependence of Large Flare Occurrence on the Magnetic Structure of Sunspots.
The Astrophysical Journal 540: 583-587. doi:10.1086/309303

\bibitem[{{Schaefer}(1989)}]{Schaefer1989}
Schaefer, B.~E. (1989)
Flashes from normal stars.
The Astrophysical Journal 337: 927-933. doi:10.1086/167162

\bibitem[{{Schaefer et al.}(2000)}]{Schaefer2000}
Schaefer, B.~E., King, J.~R., Deliyannis, C.~P. (2000)
Superflares on ordinary solar-type stars.
The Astrophysical Journal 529: 1026-1030. doi:10.1086/308325

\bibitem[{{Schrijver et al.}(2012)}]{Schrijver2012}
Schrijver, C.~J., Beer, J., Baltensperger, U., et al. (2012)
Estimating the frequency of extremely energetic solar events, based on solar, stellar, lunar, and terrestrial records.
Journal of Geophysical Research (Space Physics) 117: 8103. doi:10.1029/2012JA017706

\bibitem[{{Shibata \& Magara}(2011)}]{Shibata2011}
Shibata, K., Magara, T. (2011)
Solar Flares: Magnetohydrodynamic Processes.
Living Reviews in Solar Physics 8: 6. doi:10.12942/lrsp-2011-6

\bibitem[{{Shibata et al.}(2013)}]{Shibata2013}
Shibata, K., Isobe, H., Hillier, A., et al. (2013)
Can Superflares Occur on Our Sun?
Publications of the Astronomical Society of Japan 65: 49. doi:10.1093/pasj/65.3.49

\bibitem[{{Shibayama et al.}(2013)}]{Shibayama2013}
Shibayama, T., Maehara, H., Notsu, S., et al. (2013)
Superflares on Solar-type Stars Observed with Kepler. I. Statistical Properties of Superflares.
The Astrophysical Journal Supplement 209: 5. doi:10.1088/0067-0049/209/1/5

\bibitem[{{Shimizu}(1995)}]{Shimizu1995}
Shimizu, T. (1995)
Energetics and Occurrence Rate of Active-Region Transient Brightenings and Implications for the Heating of the Active-Region Corona.
Publications of the Astronomical Society of Japan 47: 251-263

\bibitem[{Slawson et al.}(2011)]{Slawson2011}
Slawson, R.~W., Pr{\v s}a, A., Welsh, W.~F., et al. (2011)
Kepler Eclipsing Binary Stars. II. 2165 Eclipsing Binaries in the Second Data Release
The Astronomical Journal 142: 160. doi:10.1088/0004-6256/142/5/160

\bibitem[{{Smith et al.}(2012)}]{Smith2012}
Smith, J.~C., Stumpe, M.~C., Van Cleve, J.~E., et al. (2012)
Kepler Presearch Data Conditioning II - A Bayesian Approach to Systematic Error Correction.
Publications of the Astronomical Society of the Pacific 124, 1000-1014. doi:10.1086/667697


\bibitem[{{Strassmeier}(2009)}]{Strassmeier2009}
Strassmeier, K.~G. (2009)
Starspots.
The Astronomy and Astrophysics Review 17: 251-308. doi:10.1007/s00159-009-0020-6

\bibitem[{{Stumpe et al.}(2012)}]{Stumpe2012}
Stumpe, M.~C., Smith, J.~C., Van Cleve, J.~E., et al. (2012)
Kepler Presearch Data Conditioning I - Architecture and Algorithms for Error Correction in Kepler Light Curves.
Publications of the Astronomical Society of the Pacific 124: 985-999. doi:10.1086/667698


\bibitem[{{Tsurutani et al.}(2003)}]{Tsurutani2003}
Tsurutani, B.~T., Gonzalez, W.~D., Lakhina, G.~S., \& Alex, S. (2003)
Journal of Geophysical Research (Space Physics) 108: 1268. doi:10.1029/2002JA009504

\bibitem[{{Veronig et al.}(2002)}]{Veronig2002}
Veronig, A., Temmer, M., Hanslmeier, A., et al. (2002)
Temporal aspects and frequency distributions of solar soft X-ray flares.
Astronomy and Astrophysics 382: 1070-1080. doi:10.1051/0004-6361:20011694


\bibitem[{{Welsh et al.}(2006)}]{Welsh2006}
Welsh, B.~Y., Wheatley, J., Browne, S.~E., et al. (2006)
GALEX high time-resolution ultraviolet observations of dMe flare events.
Astronomy and Astrophysics 458: 921-930. doi:10.1051/0004-6361:20065304

\bibitem[{{Wichmann et al.}(2014)}]{Wichmann2014}
Wichmann, R., Fuhrmeister, B., Wolter, U., \& Nagel, E. (2014)
Kepler super-flare stars: what are they?
Astronomy and Astrophysics 567: AA36. doi:10.1051/0004-6361/201423717

\bibitem[{{Wu et al.}(2015)}]{Wu2015}
Wu, C.-J., Ip, W.-H., \& Huang, L.-C. (2015)
The Astrophysical Journal 798: 92. doi:10.1088/0004-637X/798/2/92


\end{thebibliography}
\end{document}